# Inversion of the spin polarization of localized electrons driven by dark excitons

B. Eble[1], C. Testelin[1], F. Bernardot[1], M. Chamarro[1], and G. Karczewski[2]

[1] *Institut des NanoSciences de Paris - Universités Paris-VI et Paris-VII, CNRS UMR 7588, 4 place Jussieu, 75252 Paris cedex 05, France*

[2] *Institute of Physics, Polish Academy of Sciences, Al. Lotnikow 32/46, 02-668 Warsaw, Poland*

The creation of free excitons by absorption of circularly polarized photons, and their subsequent fast capture by donors, is at the origin of the spin polarization of donor-bound electrons. The sign of the electronic spin polarization at low density of excitation is, as expected, fixed by the helicity of the exciting light; but at high density of excitation we show that the spin polarization is of the opposite sign. This striking inversion is explained, here, by the contribution of dark excitons to mechanisms of spin polarization of localized electrons.

The spin of an individual carrier localized at the nanometer scale is a promising candidate for the implementation of spin-based devices or qubits [1], since its localization results in the suppression of spin relaxation mechanisms related to its movement or diffusion [2]. Future applications in the fields of spintronics and quantum information require fundamental studies, including investigations on generation, manipulation and readout of the electronic spin polarization. In this direction, a special attention is devoted to the all-optical approach which allows, in addition to spin polarization and readout, a very high speed of spin manipulation. Thus, the optical orientation of electronic spins is traditionally accomplished by using circularly polarized photons which transfer their angular momentum to bright excitons (excitons coupled to the light field). In semiconductors, photo-generated excitons have a limited lifetime, on the order of one nanosecond. When a very long time of preservation of the spin information is required, doped nanostructures are more appropriate since they give the opportunity to write a spin information on resident carriers. The condition for an efficient polarization of resident electrons, without external magnetic field, and under resonant excitation of charged excitons (also called trions), is that the spin relaxation time of the photo-generated holes should remain in the same order of magnitude as, or be smaller than, the lifetime of charged excitons [3, 4]. Thus, according to the selection rules, the spin polarization of the resident electrons is determined by the helicity of the circularly polarized exciting light: a spin-up ↑ (spin-down ↓) polarization is created by a σ– (σ+) exciting light. Optical techniques have been successfully used to polarize a 2D electron gas [3, 5–10], electrons localized by the Coulomb potential of donors [11–14], or localized in quantum dots (QDs) [15–20].

In this letter, we have centered our interest on electrons localized by the Coulomb potential of donors inserted at the middle of a quantum well (QW), because recent studies have shown that it is a very reproducible and homogeneous system with a very long spin coherence time [14, 21]. Immersion of donors in a QW increases the localization of the bound electron wavefunction, and the efficiency of the optical orientation of donor-bound excitons, also called $D^0X$, by splitting the light- and heavy-hole bands. The creation of polarized free excitons, and their subsequent fast capture by donors, are at the origin of the spin polarization of donor-bound electrons. We show here that, for a fixed helicity of the exciting light, the spin polarization of the resident electrons presents three regimes as a function of the excitation density: (i) at low density of excitation, the electronic spin polarization increases with the excitation density, and reaches a maximum; (ii) at intermediate densities of excitation, the spin polarization decreases until it disappears; (iii) at high densities of excitation, the spin polarization increases and saturates, but it possesses a sign opposite to the one it has at low density of excitation. We have interpreted this striking observation as an evidence of the key role played by dark excitons (excitons uncoupled to the light field) in mechanisms leading to the spin polarization of the resident electrons at high excitation density.

The studied sample consists of a CdTe/CdMgTe heterostructure grown by molecular-beam epitaxy on a (100)-oriented GaAs substrate and containing an 80 Å QW. A donor layer of iodine atoms was placed at the middle of the QW; its concentration is approximately $n_D \approx$ 2–3 $10^{10}$ cm$^{-2}$. We use a degenerate pump-probe technique, the photo-induced Faraday rotation (PFR), to generate and monitor the spin polarization of the resident electrons. A Ti:sapphire laser beam, with a 2-ps pulse duration and a repetition rate of 76 MHz, is split into the pump and probe beams. The pump beam polarization is σ+/σ– modulated at 42 kHz with an elasto-optic modulator; the probe beam is linearly polarized, and its intensity is modulated at 1 kHz with an optical chopper. After transmission through the sample, the rotation angle of the probe beam polarization is measured in an optical bridge, followed by a double lock-in amplifier. The pump power varies from about 30 µW to 10 mW, and the probe power is kept constant at 15 µW. The pump and probe beams are focused onto the sample in a spot of radius approximately equal to 20 µm.

In order to perform transmission and PFR measurements, we have chemically suppressed the GaAs substrate. Figure 1a shows the low-temperature transmission and photo-luminescence (PL) spectra. The transmission spectrum is dominated by the free-exciton absorption, peacked at 1617.5 meV. This fact arises because the oscillator strength of donor-bound excitons, even proportional to the density of donors, is much smaller than the oscillator strength of excitons. The PL spectrum, obtained at low density of excitation, is dominated by the donor-bound exciton recombination (at 1615.5 meV), indicating that a large majority of the optically created carriers relax to form preferentially donor-bound excitons.

Figure 2 shows the temporal behaviour of the low-temperature PFR signals obtained, for different excitation densities, when the pump and probe beams are tuned to the free-exciton transition. Under σ+ circularly polarized excitation, spin-polarized excitons $|+1\rangle$ are created, containing a hole with spin +3/2 and an electron with spin down (–1/2), see Fig. 1b. At low densities of excitation, the $|+1\rangle$ excitons are fastly captured by +1/2 donors, giving rise to the formation of polarized $D^0X^{+3/2}$ complexes made of two electrons in a singlet configuration and a +3/2 hole. The probe beam tuned to the free-exciton transition is then sensitive to the photo-induced difference of σ+ and σ– energies and oscillator strengths. Moreover, *via* exchange processes affecting holes or electrons [22], the free-exciton energies are sensitive to the polarization of bound excitons and bound electrons; this results in a PFR signal proportional to the bound carriers polarization. The fast decay components (within ~ 300 ps) of the PFR signals are then associated to the free- and bound-exciton spin dynamics. At low densities of excitation (< 300 µW), we note, in Fig. 2a, a nonzero PFR signal at negative pump-probe delays, indicating that the donor spin polarization is not fully relaxed within the 13-ns repetition period of the laser pulses. In the following we will address the behaviour of the long-time component of the PFR signals, which is unambiguously associated to the net spin polarization of the donor-bound electrons, the only species present in the sample when the recombination of free excitons (on the order of tens of picoseconds [6, 23]), the capture of free excitons by donors, and the recombination of $D^0X$ complexes (on the order of several hundreds of picoseconds [24]) are finished.

Figures 2b–f show that, as the density of excitation increases, the amplitude of the long-living PFR signal decreases, goes to zero, and is finally reversed becoming clearly negative [25]; a transient signal of several hundreds of picoseconds accompanies this reversal. To understand this behaviour and the whole dynamics of the free excitons, the bound excitons and the bound electrons, we have used rate equations taking into account eight different populations (see Fig. 1b). After a resonant σ+ excitation of the free-exciton transition, a population of $|+1\rangle$ excitons is created, with a radiative lifetime denoted $\tau_r$. From this initial population, by spin flip of electrons (holes), $|+2\rangle$ ($|-2\rangle$) dark excitons can be generated; by flipping the whole exciton spin, mutual changes between $|+1\rangle$ and $|-1\rangle$, or between $|+2\rangle$ and $|-2\rangle$, are allowed. The corresponding involved characteristic times $\tau_e$, $\tau_h$ and $\tau_{exch}$ are mentioned in Fig. 1b. Because of the bright-dark energy splitting (Δ ~ 200 μeV [6]), we use different times for spin flips from bright to dark excitons: $\tau_e^{down} = \tau_e$, $\tau_h^{down} = \tau_h$, and for spin flips from dark to bright excitons: $\tau_e^{up} = \tau_e \exp(\Delta/kT) \approx 3\tau_e$, $\tau_h^{up} = 3\tau_h$ (T = 2 K). Neutral donors containing an electron with spin up, $D_\uparrow^0$, can capture $|+1\rangle$ ($|-2\rangle$) excitons to form bound exciton complexes $D^0X^{+3/2}$ ($D^0X^{-3/2}$), meanwhile neutral donors containing an electron with spin down, $D_\downarrow^0$, can capture $|-1\rangle$ ($|+2\rangle$) excitons to form donor-bound excitons $D^0X^{-3/2}$ ($D^0X^{+3/2}$). These capture processes are characterized by a capture coefficient γ, and play a key role in the polarization mechanisms by transforming *dark* excitons $|\pm2\rangle$ into *bright* donor-bound excitons $D^0X^{\pm3/2}$. $T_h$ is the spin relaxation time of the donor-bound exciton, and $T_R$ its lifetime. The recombination of a $D^0X^{+3/2}$ ($D^0X^{-3/2}$) complex leads to a spin-up (spin-down) bound electron. The spin relaxation time of the donor-bound electrons is denoted $T_e$. The dashed lines in Fig. 2 are the results of this model – the fitting parameters are given in the figure caption; the calculated donor-bound spin polarization $[n_\downarrow(t) - n_\uparrow(t)]/n_D$ is scaled to the PFR signals with a single constant factor [26]. We have taken into account the saturation of excitons generation when the pump power, P, increases [27]. The concentration of created $|+1\rangle$ excitons at t = 0, $n_{+1}^0$, writes:

$$n_{+1}^0 = A_\infty \left[\frac{P}{P_s}\right] / \left[1 + \frac{P}{P_s}\right], \qquad (1)$$

where $A_\infty$ and $P_s$ are two fitting parameters.

Figure 3a shows the PFR signals (red points) obtained at a pump-probe delay equal to 13 ns, for different pump densities. The solid lines in the same figure represent the theoretical polarization of the bound electrons, for several $\tau_e$ values, the other fitting parameters being kept constant. The different observed regimes are well described by the already mentioned rate equations and fitting parameters.

At low density of excitation, when the pump creates a population of free excitons which is much smaller than the half of the total population of donors, $n_{+1}^0 \ll n_D/2$, the probability for the free excitons to become bound to donors is very high, and then the electronic polarization increases linearly. A saturation of the electronic polarization is reached for $n_{+1}^0 \approx n_D/2$. If the pump density is further increased in order to initially obtain a higher population of free excitons ($n_D/2 < n_{+1}^0 < n_D$), the electronic polarization decreases and goes to zero. For the highest densities of excitation, $n_{+1}^0 > n_D$, the spin polarization of the donor-bound electrons is inverted. The maximum value of the electronic polarization, its decrease and its inversion are the results of a complicated function of the different rate parameters governing the free exciton populations, their capture by donors, and also the recombination and relaxation rates of the donor-bound excitons. To get a more precise idea of the role played by the different populations and parameters in the construction of the inverted electronic polarization, we have split the processes into two steps: first, the (rapid) relaxation and trapping of the free excitons, and second, the (slow) relaxation and recombination of the donor-bound excitons. Figure 3b illustrates the first step; it shows the integrated populations of the different free excitons which are captured by neutral donors, as a function of the pump power density:

$$n^*_{+1,-2(-1,+2)} = \gamma \int_0^\infty n_{+1,-2(-1,+2)}(t) n_{\uparrow(\downarrow)}(t)\, dt. \quad (2)$$

We note that, whatever the density of excitation, the $|+1\rangle$ excitons are captured by neutral donors in a preferential manner, contributing by $n^*_{+1}$ to the initial population of $D^0X^{+3/2}$. At high excitation densities, the population of captured $|+2\rangle$ dark excitons, $n^*_{+2}$, becomes the second more important population, while, at low excitation densities, it is comparable to the captured $|-1\rangle$ and $|-2\rangle$ populations, denoted respectively $n^*_{-1}$ and $n^*_{-2}$. Once this first step accomplished, a linear system of four rate equations, containing the four populations of donor-bound excitons and electrons, gives the electronic polarization:

$$\frac{n_\downarrow - n_\uparrow}{n_D} = n_D^{-1}(e^{t_{obs}/T_e} - 1)^{-1} \left\{ [(1-a)n^*_{+1} + (1+a)n^*_{-2}] - [(1-a)n^*_{-1} + (1+a)n^*_{+2}] \right\}, \quad (3)$$

where $a = (1 + T_R/T_h)^{-1}$ and $t_{obs} = 13$ ns. Equation (3) shows that $n^*_{+1}$ and $n^*_{-2}$ contribute positively to $n_\downarrow - n_\uparrow$, and $n^*_{-1}$ and $n^*_{+2}$ contribute negatively. Figure 3c shows the evolution, as a function of the pump density, of the positive and negative contributions to the electronic polarization in our sample; their crossing at ~ 1000 µW explains the reversing of the spin polarization of the donor-bound electrons.

Equation (3) shows that, at low densities of excitation, the contributions of $n^*_{+2}$ and $n^*_{-2}$ cancel each other, and the polarization created by $n^*_{+1}$ is, in general, larger than the polarization created by $n^*_{-1}$, then $n_\downarrow - n_\uparrow > 0$. At high density of excitation, three cases appear clearly. In the first case, $T_h \gg T_R$ and then $a \rightarrow 1$: only the trapped populations of dark excitons $|+2\rangle$ and $|-2\rangle$ contribute to the net electronic polarization, which becomes negative if $n^*_{-2} < n^*_{+2}$. In the second case, which is the case of our sample, $T_h \approx T_R$ and $a \approx 1/2$: the inversion is possible if $3(n^*_{+2} - n^*_{-2}) > n^*_{+1} - n^*_{-1}$. The factor of 3 arises from the mechanisms by which the trapped $|+2\rangle$ and $|+1\rangle$ contribute to the spin polarization of the donor-bound electrons: once trapped on a $D^0_\downarrow$, the $|+2\rangle$ creates a $D^0X^{+3/2}$ which only needs to recombine to create a $D^0_\uparrow$; this negative process is more efficient than the polarization mechanism due to the trapping of a $|+1\rangle$ exciton on a $D^0_\uparrow$, because the created $D^0X^{+3/2}$ then needs to spin-flip before recombination to release a $D^0_\downarrow$. In the third case, $T_h \ll T_R$ and $a \rightarrow 0$: the inversion is possible if $n^*_{+2} - n^*_{-2} > n^*_{+1} - n^*_{-1}$, a condition which is not realized for a reasonable set of rates governing the different populations of free excitons. Free trions in a n-doped QW are a good example of this third case. Recently a non-monotonous behaviour has been observed in samples containing a 2D electron gas in a CdTe QW, but no inversion of the polarization of the 2D electron gas was observed [10]. A crucial condition to observe a negative polarization is that $T_h \approx T_R$. This condition is satisfied when the resident electron is localized, which is the case of electrons bound to donors or electrons trapped in small QDs.

It has to be mentioned that a negative rate of circular polarization of PL, associated to the polarization of resident electrons, has been recently observed in n-doped QDs under an excitation at the wetting-layer energy. It has been explained to be the result of an interplay between periodic electron-hole flip flops, due to the anisotropic exchange interaction, and the thermalization of Pauli-blocked electron-pair configurations [28]. We underline that the mechanisms and consequences of the optical orientation of resident electrons by a non-resonant light are very different for electrons bound to donors or for electrons localized in QDs. First, contrary to n-doped QDs, the rate of circular polarization of the $D^0X$ PL is always positive, even under inversion of the electronic polarization, as can be deduced from Fig. 3b and expression $[(n^*_{+1} + n^*_{+2}) - (n^*_{-1} + n^*_{-2})]/[\Sigma n^*]$ for this PL polarization rate. Second, unlike the spin polarization of electrons localized on donors, the spin polarization of electrons confined in QDs is always determined by the helicity of the exciting light.

In conclusion, when the creation of free excitons is used to polarize resident electrons localized by the Coulomb potential of donors, the sign of this polarization can be governed by the pump density of excitation: at low density, it is opposite to the helicity of the pump beam, and at high densities it is inverted. This experimental observation demonstrates an important participation, until now neglected, of the free dark excitons in the mechanisms of the electronic spin polarization. This result opens doors for manipulations of the electronic spin *via* dark excitons.

The authors acknowledge financial support of the Ile-de-France Regional Council through the "projet SESAME 2003" n°E.1751 and C-Nano post-doctoral fellowship 2007.

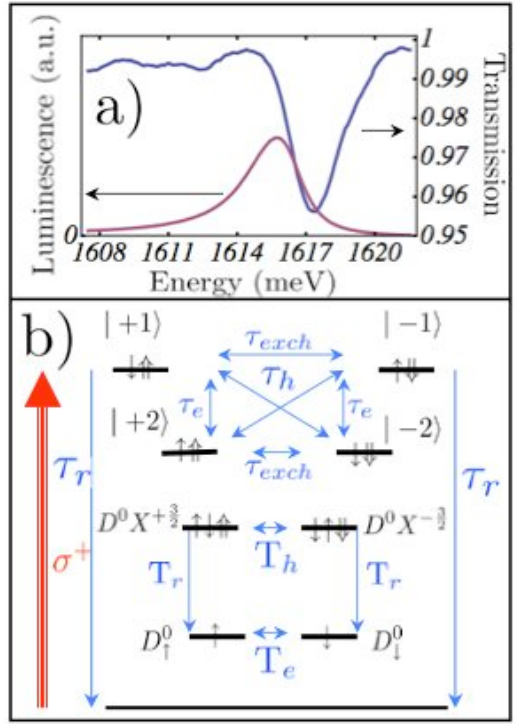

**Figure 1** a) Low-temperature PL (red) and transmission (blue) spectra of a single CdTe/CdMgTe QW containing a plane of donors ($n_D$ = 2–3 $10^{10}$ cm$^{-2}$). b) Scheme of the energy levels involved in the mechanism of polarization of the electrons bound to donors.

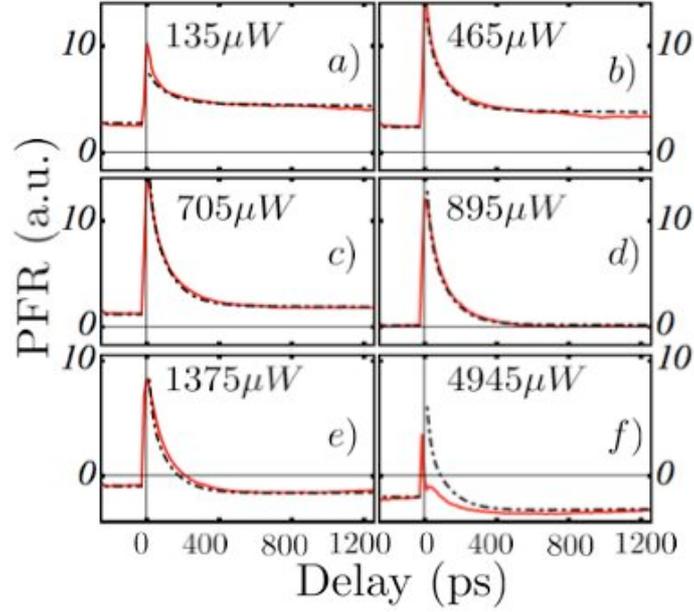

**Figure 2** The red continuous lines represent experimental low-temperature PFR signals, and the black dashed lines represent calculated (and scaled) $[n_\downarrow(t) - n_\uparrow(t)]/n_D$, for different pump excitation densities: a) 135 µW; b) 465 µW; c) 705 µW; d) 895 µW; e) 1375 µW; f) 4945 µW. The values of the different characteristic times used in the eight non-linear rate equations, relating the populations of the species shown in Fig. 1b, are the following: $\tau_r$ = 50 ps, $\tau_e$ = 25 ps, $\tau_h$ = 15 ps, $\tau_{exch}$ = 8 ps, $T_R$ = 190 ps, $T_h$ = 230 ps, $T_e$ = 25 ns; also: $(\gamma n_D)^{-1}$ = 5 ps is the capture time of free excitons by donors. The initial $|+1\rangle$ population is given by Eq. (1), with $A_\infty/n_D$ = 1.5 and $P_s$ = 550 µW.

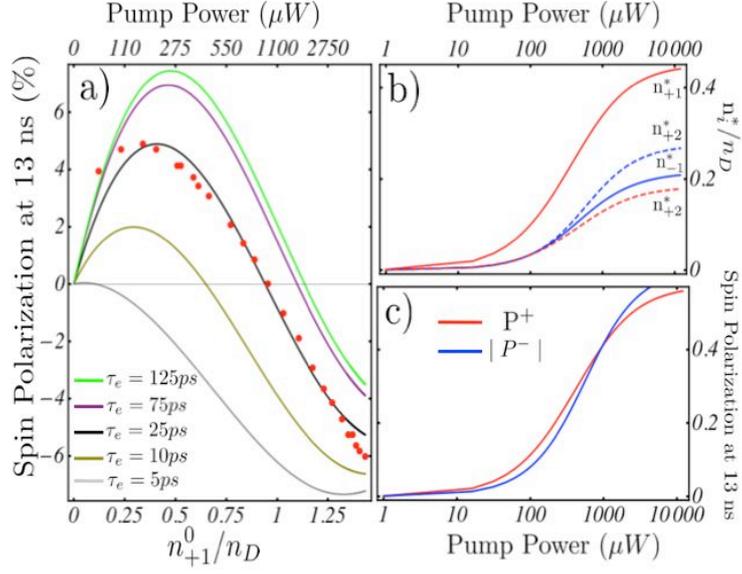

**Figure 3** a) The red points represent PFR signals obtained at 13 ns, as a function of the pump power. The solid lines are the calculated spin polarization $[n_\downarrow(13ns) - n_\uparrow(13ns)]/n_D$, as a function of the pump power for different $\tau_e$ values. b) Integrated free exciton populations trapped on the donors (Eq. (2)), normalized to the nominal concentration of donors, as a function of pump power. c) Positive and negative contributions to the spin polarization at 13 ns, *versus* pump power; the difference of both curves gives the spin polarization of the donor-bound electrons (here, $P^+ = n_D^{-1}(e^{t_{obs}/T_e} - 1)^{-1}[(1-a)n^*_{+1} + (1+a)n^*_{-2}]$ and $P^- = -n_D^{-1}(e^{t_{obs}/T_e} - 1)^{-1}[(1-a)n^*_{-1} + (1+a)n^*_{+2}]$, see Eq. (3)).

BIBLIOGRAPHY

[1] *Semiconductor Spintronics and Quantum Computing*, edited by D. D. Awschalom, D. Loss and N. Samarth (Springer, New York, 2002).

[2] A. V. Khaetskii and Y. V. Nazarov, Phys. Rev. B **61**, 12639 (2000); A. V. Khaetskii and Y. V. Nazarov, Phys. Rev. B **64**, 125316 (2001).

[3] J. Tribollet, F. Bernardot, M. Menant, G. Karczewski, C. Testelin, and M. Chamarro, Phys. Rev. B **68**, 235316 (2003).

[4] M. Chamarro, F. Bernardot, and C. Testelin, J. Phys.: Condens. Matter **19**, 445007 (2007).

[5] J. M. Kikkawa, I. P. Smorchkova, N. Samarth, and D. D. Awschalom, Science **277**, 1284 (1997).

[6] E. Vanelle, M. Paillard, X. Marie, T. Amand, P. Gilliot, D. Brinkmann, R. Lévy, J. Cibert, and S. Tatarenko, Phys. Rev. B **62**, 2696 (2000).

[7] T. A. Kennedy, A. Shabaev, M. Scheibner, Al. L. Efros, A. S. Bracker, and D. Gammon, Phys. Rev. B **73**, 045307 (2006).

[8] I. Ya. Gerlovin, Yu. P. Efimov, Yu. K. Dolgikh, S. A. Eliseev, V. V. Ovsyankin, V. V. Petrov, R. V. Cherbunin, I. V. Ignatiev, I. A. Yugova, L. V. Fokina, A. Greilich, D. R. Yakovlev, and M. Bayer, Phys. Rev. B **75**, 115330 (2007).

[9] Z. Chen, R. Bratschitsch, S. G. Carter, S. T. Cundiff, D. R. Yakovlev, G. Karczewski, T. Wojtowicz, and J. Kossut, Phys. Rev. B **75**, 115320 (2007).

[10] E. A. Zhukov, D. R. Yakovlev, M. Bayer, M. M. Glazov, E. L. Ivchenko, G. Karczewski, T. Wojtowicz, and J. Kossut, Phys. Rev. B **76**, 205310 (2007).

[11] *Optical Orientation*, edited by F. Meier and B. Zakharchenya, Modern Problems in Condensed Matter Sciences Vol. 8 (North-Holland, Amsterdam, 1984).

[12] J. M. Kikkawa and D. D. Awschalom, Phys. Rev. Lett. **80**, 4313 (1998).

[13] R. I. Dzhioev, K. V. Kavokin, V. L. Korenev, M. V. Lazarev, B. Ya. Meltser, M. N. Stepanova, B. P. Zakharchenya, D. Gammon, and D. S. Katzer, Phys. Rev. B **66**, 245204 (2002).

[14] J. Tribollet, E. Aubry, G. Karczewski, B. Sermage, F. Bernardot, C. Testelin, and M. Chamarro, Phys. Rev. B **75**, 205304 (2007).

[15] S. Cortez, O. Krebs, S. Laurent, M. Senes, X. Marie, P. Voisin, R. Ferreira, G. Bastard, J.-M. Gérard, and T. Amand, Phys. Rev. Lett. **89**, 207401 (2002).

[16] P.-F. Braun, X. Marie, L. Lombez, B. Urbaszek, T. Amand, P. Renucci, V. K. Kalevich, K. V. Kavokin, O. Krebs, P. Voisin, and Y. Masumoto. Phys. Rev. Lett. **94**, 116601 (2005).

[17] M. V. Gurudev Dutt, Jun Cheng, Bo Li, Xiaodong Xu, Xiaoqin Li, P. R. Berman, D. G. Steel, A. S. Bracker, D. Gammon, Sophia E. Economou, Ren-Bao Liu, and L. J. Sham, Phys. Rev. Lett. **94**, 227403 (2005).

[18] A. Greilich, R. Oulton, E. A. Zhukov, I. A. Yugova, D. R. Yakovlev, M. Bayer, A. Shabaev, Al. L. Efros, I. A. Merkulov, V. Stavarache, D. Reuter, and A. Wieck, Phys. Rev. Lett. **96**, 227401 (2006).

[19] I. A. Akimov, D. H. Feng, and F. Henneberger, Phys. Rev. Lett. **97**, 056602 (2006).

[20] E. Aubry, C. Testelin, F. Bernardot, M. Chamarro and A. Lemaître, Appl. Phys. Lett. **90**, 242113 (2007).

[21] Kai-Mei C. Fu, Wenzheng Yeo, Susan Clark, Charles Santori, Colin Stanley, M. C. Holland, and Yoshihisa Yamamoto, Phys. Rev. B **74**, 121304 (2006).

[22] M. Combescot, J. Tribollet, G. Karczewski, F. Bernardot, C. Testelin, and M. Chamarro, Europhys. Lett. **71**, 431 (2005).

[23] A. Vinattieri, Jagdeep Shah, T. C. Damen, D. S. Kim, L. N. Pfeiffer, M. Z. Maialle, and L. J. Sham, Phys. Rev. B **50**, 10868 (1994).

[24] S. Seto, K. Suzuki, M. Adachi, K. Inabe, Physica B **302-303**, 307 (2001).

[25] This effect has also been observed in other QWs containing donors with concentrations in the range $10^{10}$–$10^{11}$ cm$^{-2}$.

[26] The good fitting appearing in Figs. 2a–e indicates that the spin polarization of donor-bound electrons is the dominant contribution to the PFR signal.

[27] S. Schmitt-Rink , D. S. Chemla, and D. A. B. Miller, Phys. Rev. B **32**, 6601 (1985); Daming Huang, Jen-Inn Chyi, and Hadis Morkoc, Phys. Rev. B **42**, 5147 (1990).

[28] S. Laurent, M. Senes, O. Krebs, V. K. Kalevich, B. Urbaszek, X. Marie, T. Amand, and P. Voisin, Phys. Rev. B **73**, 235302 (2006).